\documentclass[aps,prl,groupedaddress,preprint,showpacs]{revtex4}

\usepackage[dvips]{graphicx}

\begin{document}

\title{Generation and characterization of a source of wavelength
division multiplexing quantum key distribution}

\author{Atsushi Yabushita}
 \author{Takayoshi Kobayashi}
 \affiliation{Core Research for Evolutional Science and
 Technology(CREST)}
\affiliation{Japan Science and Technology Corporation (JST)}
\affiliation{Department of Physics, Graduate School of Science,
 University of Tokyo, 7-3-1 Hongo, Bunkyo, Tokyo, 113-0033, Japan}

\date{\today}

\begin{abstract}
 Using spontaneous parametric down-conversion, photon pairs entangled in
 frequency and polarization were generated. After frequency resolving the
 photon pairs, the polarization correlations were measured on several
 polarization basis, and it was confirmed that the frequency resolved
 photon pairs were entangled in polarization, indicating the photon
 pairs can be used as a source of wavelength division multiplexing
 quantum key distribution.
\end{abstract}


\maketitle

\section{Introduction}

Photon pairs, generated by spontaneous parametric
down-conversion (SPDC), are entangled in various parameters. The
polarization entanglement of these SPDC photon pairs has been used in
varieties of quantum experiments to demonstrate quantum teleportation,
quantum key distribution, Bell-inequality violations, and
others~\cite{quantum-review}.

The wave-vector entanglement has also been used in various experiments, like
quantum imaging~\cite{quantum-imaging-lithography,quantum-imaging},
photonic de Broglie wavelength
measurement~\cite{deBroglie,deBroglie2,deBroglie3}, quantum
interference~\cite{quantum-interference2,quantum-interference3}, and
quantum
lithography~\cite{quantum-lithography,quantum-lithography2,quantum-lithography3}.
In case of quantum imaging, the information about the shape of a spatial
filter is transferred by entangled photon pairs. The wave vector is a
continuous parameter, therefore its entanglement can send much more
information than the polarization entanglement. Actually, a protocol for
quantum key distribution was proposed based on a system whose dimension
is higher than 2 in Ref.~\cite{highdim-hilbert-crypt-th}.

Photon pairs entangled in frequency were used for nonlocal pulse
shaping~\cite{freq-entangle-nonlocal-pulseshaping} and
spectroscopy~\cite{yab-SPDC-spectroscopy}. In
Ref.~\cite{yab-SPDC-spectroscopy}, signal and idler photons of SPDC
photon pairs were separated from each other by a polarizing beam
splitter, which destroys the polarization entanglement.
Replacing the polarizing beam splitter with a non-polarizing beam
splitter, the photon pairs are got to be entangled not only in frequency
but also in polarization. In this case, the photon pairs are entangled
in polarization even after they are resolved by their frequency,
therefore they can be used as a source of wavelength-division
multiplexing quantum key distribution (WDM-QKD).

In this article, the photon pairs entangled polarization even after
being resolved by their frequency were generated and their
characterization was done. The result of the measurement shows that
they are applicable as a source of WDM-QKD.

\section{Experiment}

   \begin{figure}[htbp]
    \begin{center}
     \includegraphics[scale=0.4]{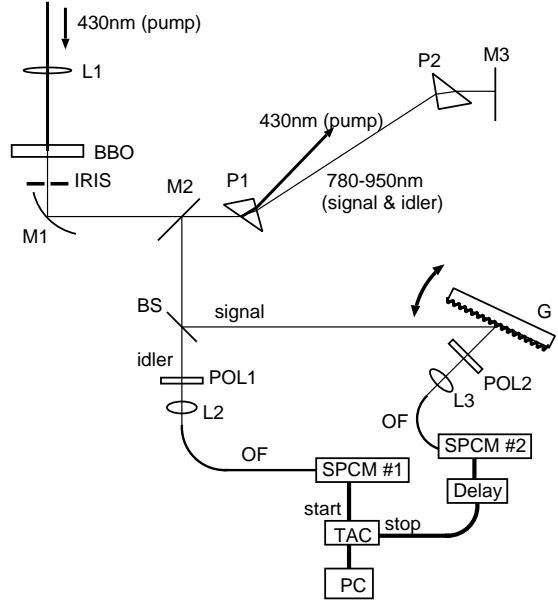}
    \end{center}
    \caption{Experimental setup.
% See the text for a detailed description.
    \label{figures-setup}}
   \end{figure}
The schematic drawing of our experimental setup is shown in
Fig.~\ref{figures-setup}.
Frequency-nondegenerate photon pairs were generated by spontaneous
parametric down-conversion (SPDC) in a
1-mm-thick type II BBO ($\beta$-BaB$_2$O$_4$) crystal pumped by the
second harmonic light at 859.4~nm (1.5~mW) of a cw Ti:sapphire
laser. The signal and idler photons of the SPDC pairs were
emitted conically from the focal waist where the pump light was focused on the
BBO crystal by a lens of 1000-mm focal length (L1). An iris diaphragms
(IRIS) was placed at the 100~mm behind the BBO crystal in order to
select spatially the crossing point of the signal light cone and the idler light
cone where the SPDC photon pairs are entangled. These pairs passed
through the iris were collimated by an off-axis parabolic mirror (M) of
25.4-mm focal length. A prism (P1) was used to eliminate the remainder of
the pump light, which can be a noise source in the experiments. Another prism
(P2) was used to compensate frequency dependent angular dispersion
induced by the prism P1. When the light beam passes through the prism
pairs, the beam height was lowered by a mirror (M3), and only the SPDC
pairs were picked out by another mirror (M2). The signal and idler
photons were separated from each other by a non-polarizing beam splitter
(BS).
A linear polarizer (POL1) was placed in the light path of idler photons
(photons transmitted the BS). The polarizer was on a motorized continuous
rotation stage, and its angle was computer-controlled. Signal photons
(photons reflected by the BS) were diffracted by a grating (G) (1400
grooves/mm). Another linear polarizer (POL2) was placed in the signal
light path, and its angle was also computer-controlled.

\section{Theory}
\label{section-theory}

For simplification, the state of the SPDC photon pair of signal ($j=s$)
and idler ($j=i$) photons is assumed to be a pure biphoton state in this
article as follows
\begin{eqnarray}
|\Psi\rangle = \frac{1}{\sqrt{1+f^2}}\left(|H\rangle_s|V\rangle_i+fe^{i\alpha}|V\rangle_s|H\rangle_i\right),
\label{eq-biphoton}
\end{eqnarray}
where $|H\rangle_j$ and $|V\rangle_j$ are single-photon states with
horizontal and vertical polarizations, respectively.
The electric fields which transmit the polarizer POL1 and POL2 are
expressed as
\begin{eqnarray}
 E_i^{(+)} & = & a_{i,H} \sin\theta_i + a_{i,V} \cos\theta_i \nonumber \\
 E_s^{(+)} & = & a_{s,H} \sin\theta_s + a_{s,V} \cos\theta_s.
\label{eq-field}
\end{eqnarray}
Here $a_{j,H}$ and $a_{j,V}$ are annihilation operators of horizontally
polarized photons and vertically polarized photons, respectively, and
$\theta_j$ is the angle between the polarizer axis and the vertical
axis.

\begin{figure}[htbp]
    \begin{center}
     \includegraphics[scale=0.18]{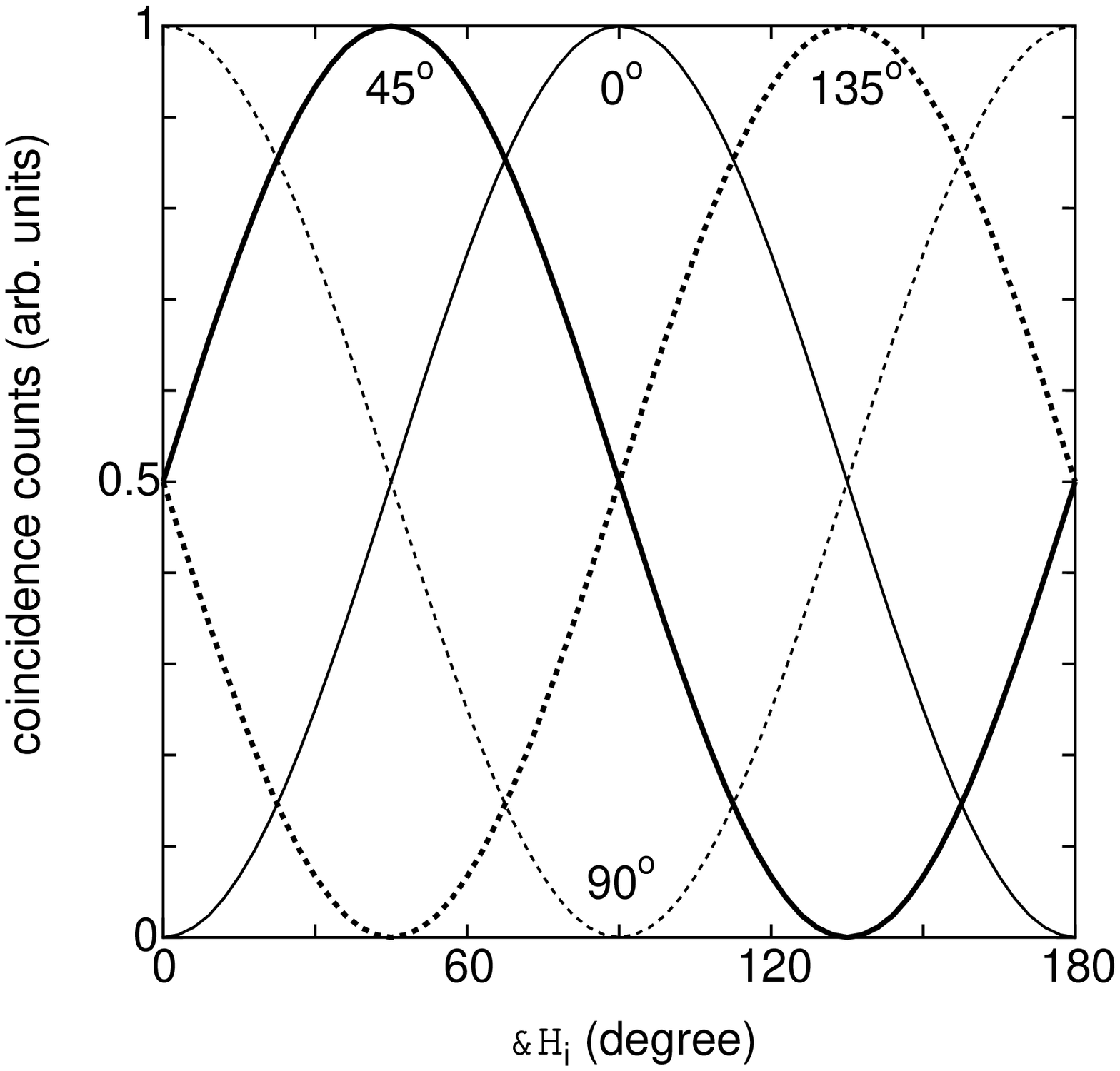}
(a)
     \includegraphics[scale=0.18]{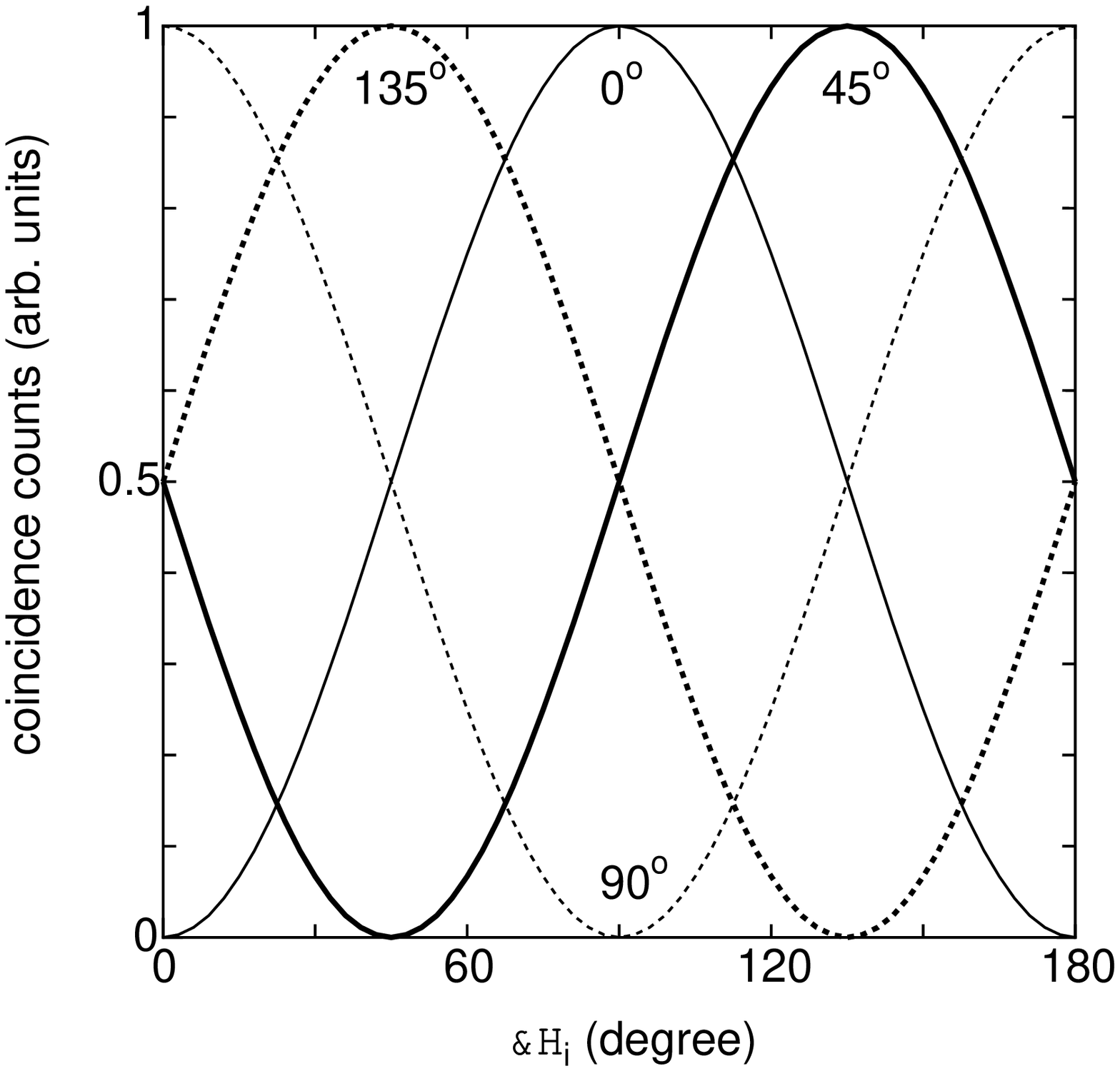}
(b)
     \includegraphics[scale=0.18]{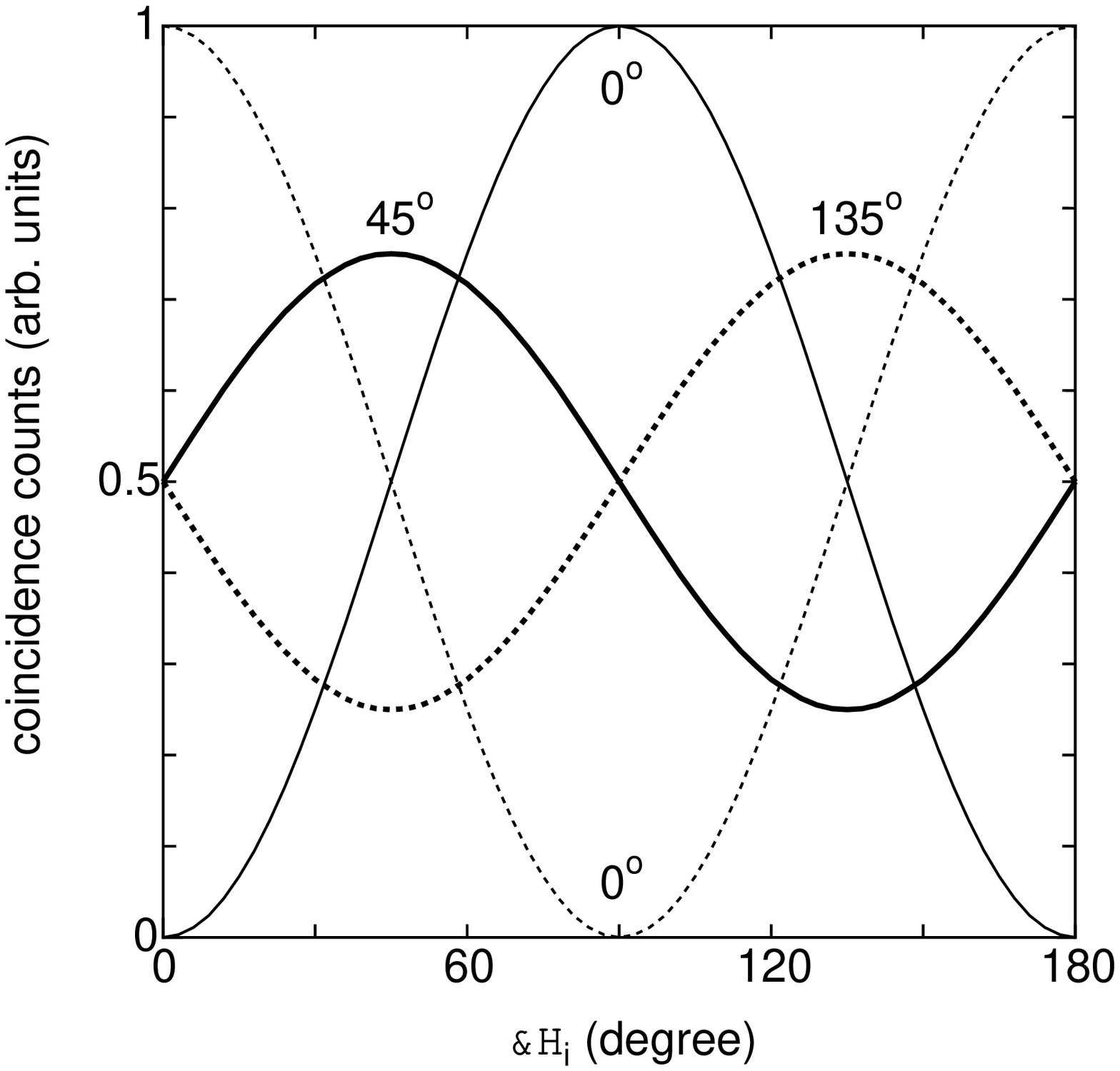}
(c)
     \includegraphics[scale=0.18]{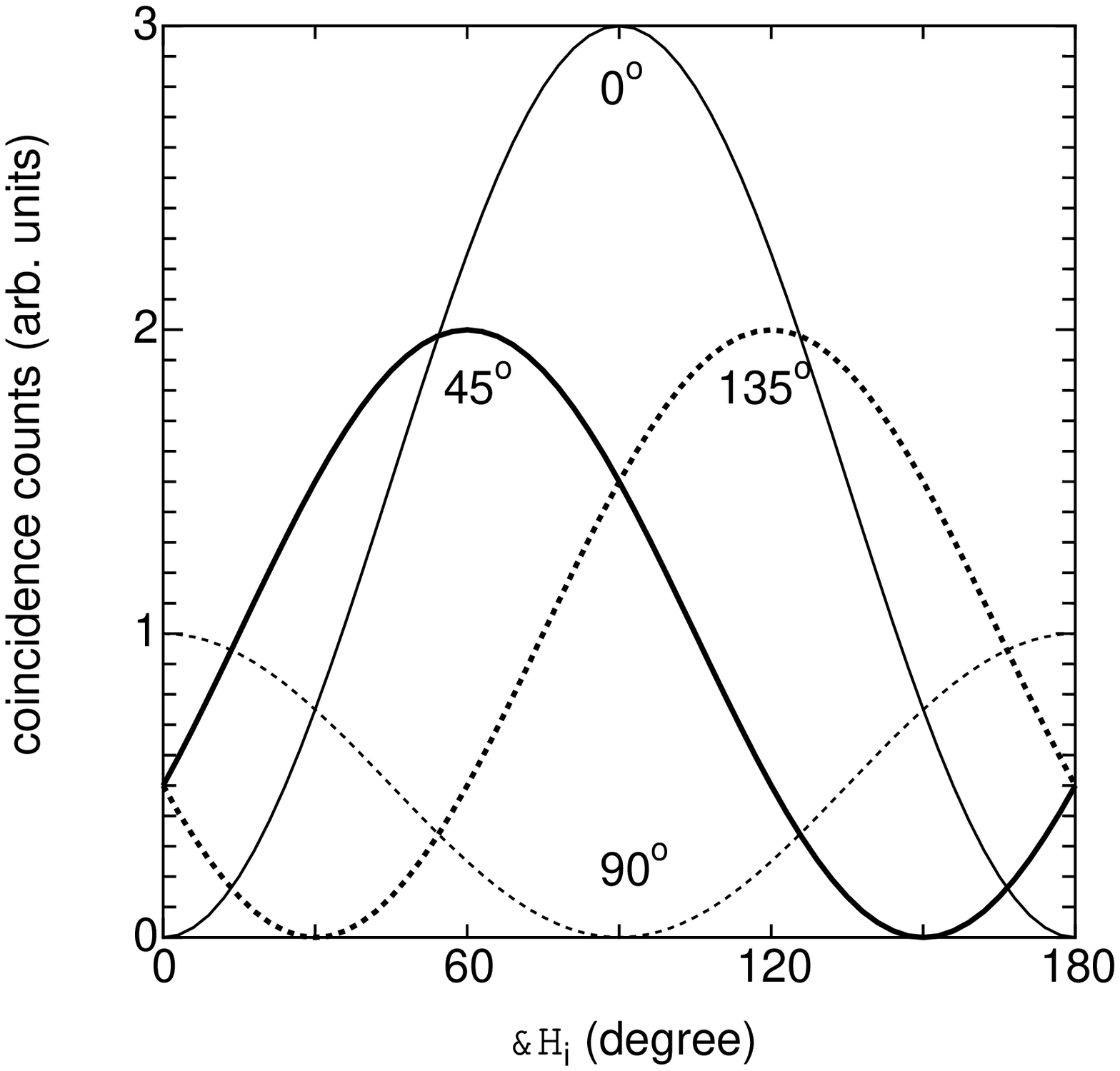}
(d)
    \end{center}
    \caption{Simulated results of the polarization correlations using of
  Eq.~(\ref{eq-coinc-rate-2}) for several configurations. (a)
    $f=1,\alpha=0^o$. (b) $f=1,\alpha=180^o$. (c) $f=1,\alpha=60^o$. (d)
    $f=1.73,\alpha=0^o$. \label{data-simulation}}
   \end{figure}

The coincidence counting rate of SPDC photon pairs
$R_c(\alpha,f,\theta_s,\theta_i)$ can be written as
\begin{eqnarray}
R_c(\alpha,f,\theta_s,\theta_i) & \propto & \langle\psi|E_s^{(-)}E_i^{(-)}E_s^{(+)}E_i^{(+)}|\psi\rangle \nonumber \\
& = & \left|\langle0|E_s^{(+)}E_i^{(+)}|\psi\rangle\right|^2.
\label{eq-coinc-rate}
\end{eqnarray}
Substituting Eqs.~(\ref{eq-biphoton})~(\ref{eq-field}) into
Eq.~(\ref{eq-coinc-rate}), we get
\begin{eqnarray}
 R_c(\alpha,f,\theta_s,\theta_i) & \propto & \frac{1+f^2+2f\cos\alpha}{4}\sin^2{(\theta_s+\theta_i)} \nonumber \\
& & +\frac{1+f^2-2f\cos\alpha}{4}\sin^2{(\theta_s-\theta_i)} \nonumber \\
& & +\frac{1-f^2}{2}\sin(\theta_s+\theta_i)\sin(\theta_s-\theta_i)
\label{eq-coinc-rate-2}
\end{eqnarray}
Results calculated with sets of parameters are shown in
Fig.~\ref{data-simulation}.

In this article, $\Theta_i(\theta_s)$ is defined as the angle
$\theta_i$ at which the coincidence counting rate is maximized. When
$f=1$, Fig.~\ref{data-simulation} shows $\Theta_i(45^o)$ and
$\Theta_i(135^o)$ are rotated by $45^o$ clockwise and anti-clockwise,
respectively, from $\Theta_i(0^o)$. However, when $f\neq1$, the angular
rotation is smaller than $45^o$. The shift is calculated to be $30^o$ in
case of $f=1.73$.

In case of $\alpha=0^o$ or $180^o$, the visibility of the polarization
correlations in Fig.\ref{data-simulation} is unity. However it is smaller
than one in other cases, like $\alpha=60^o$. In case of $f=1$ and
$\alpha=0^o,180^o$, the state is maximally entangled, and it is one of
the Bell states. In the next section, the polarization correlations
between signal photons and idler photons are measured under several
conditions and the polarization entanglement of the SPDC photon pairs
are measured.

When the state of the SPDC photon pair is a mixed state as follows
\begin{eqnarray}
 |\Psi\rangle =(|H\rangle_s+|V\rangle_s)\cdot(|H\rangle_i+|V\rangle_i),
\label{eq-mixed}
\end{eqnarray}
the coincidence counting rate of SPDC photon pairs
$R_c(\theta_s,\theta_i)$ can be calculated
as follows
\begin{eqnarray}
 R_c(\theta_s,\theta_i) \propto \sin^2(\theta_i+45^o)\sin^2(\theta_s+45^o),
\label{eq-coinc-rate-mixed}
\end{eqnarray}
by substituting Eq.~(\ref{eq-mixed}) into Eq.~(\ref{eq-coinc-rate}).
It
%Equation~(\ref{eq-coinc-rate-mixed})
shows
$\Theta_i(\theta_s)$, the angle
$\theta_i$ at which the coincidence counting rate is maximized,
is independent of $\theta_s$,
and $\Theta_i(0^o)=\Theta_i(45^o)=\Theta_i(135^o)$.

\section{Results and Discussion}

\begin{figure}[htbp]
    \begin{center}
     \includegraphics[scale=0.3]{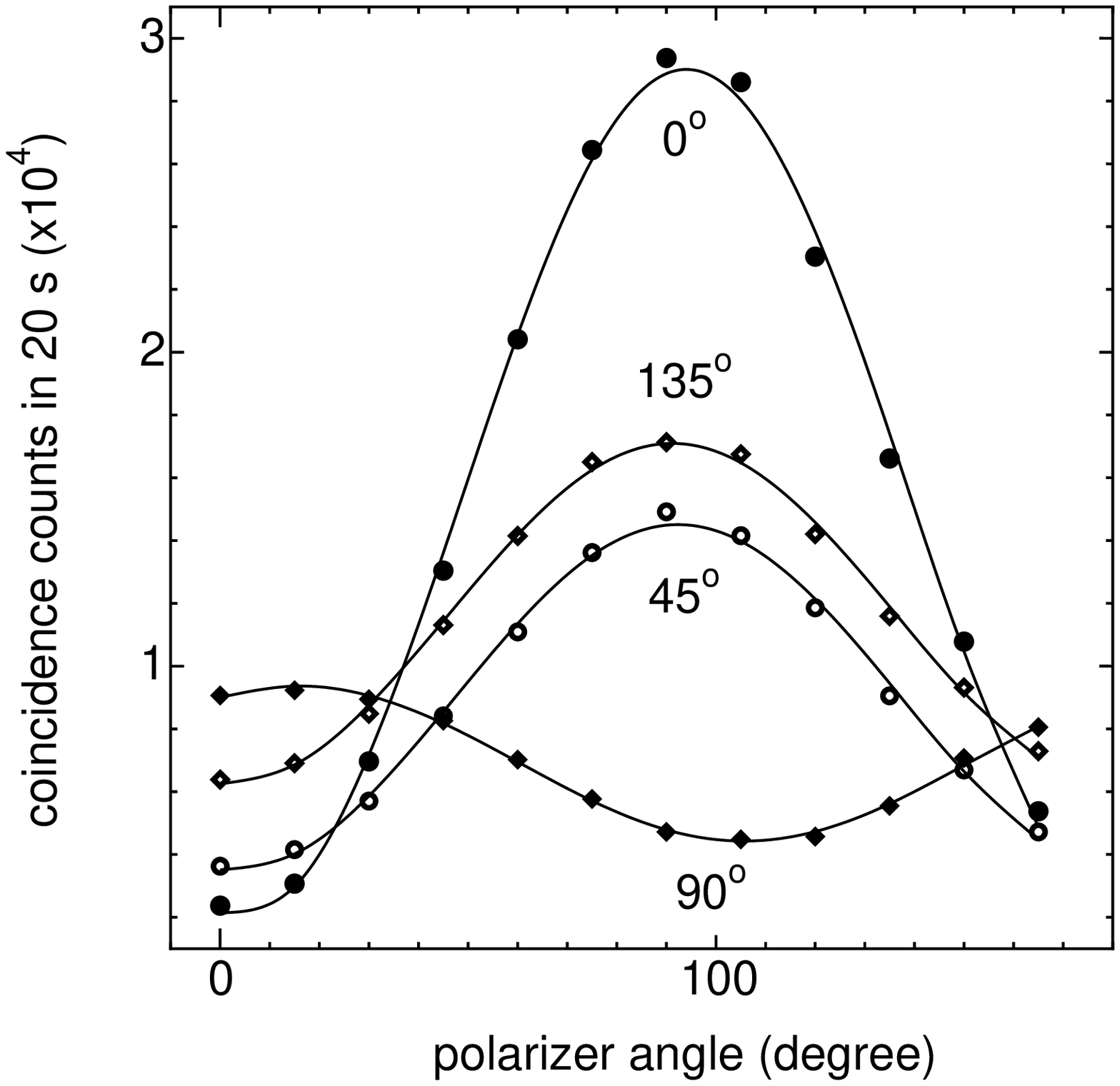}
(a)
     \includegraphics[scale=0.3]{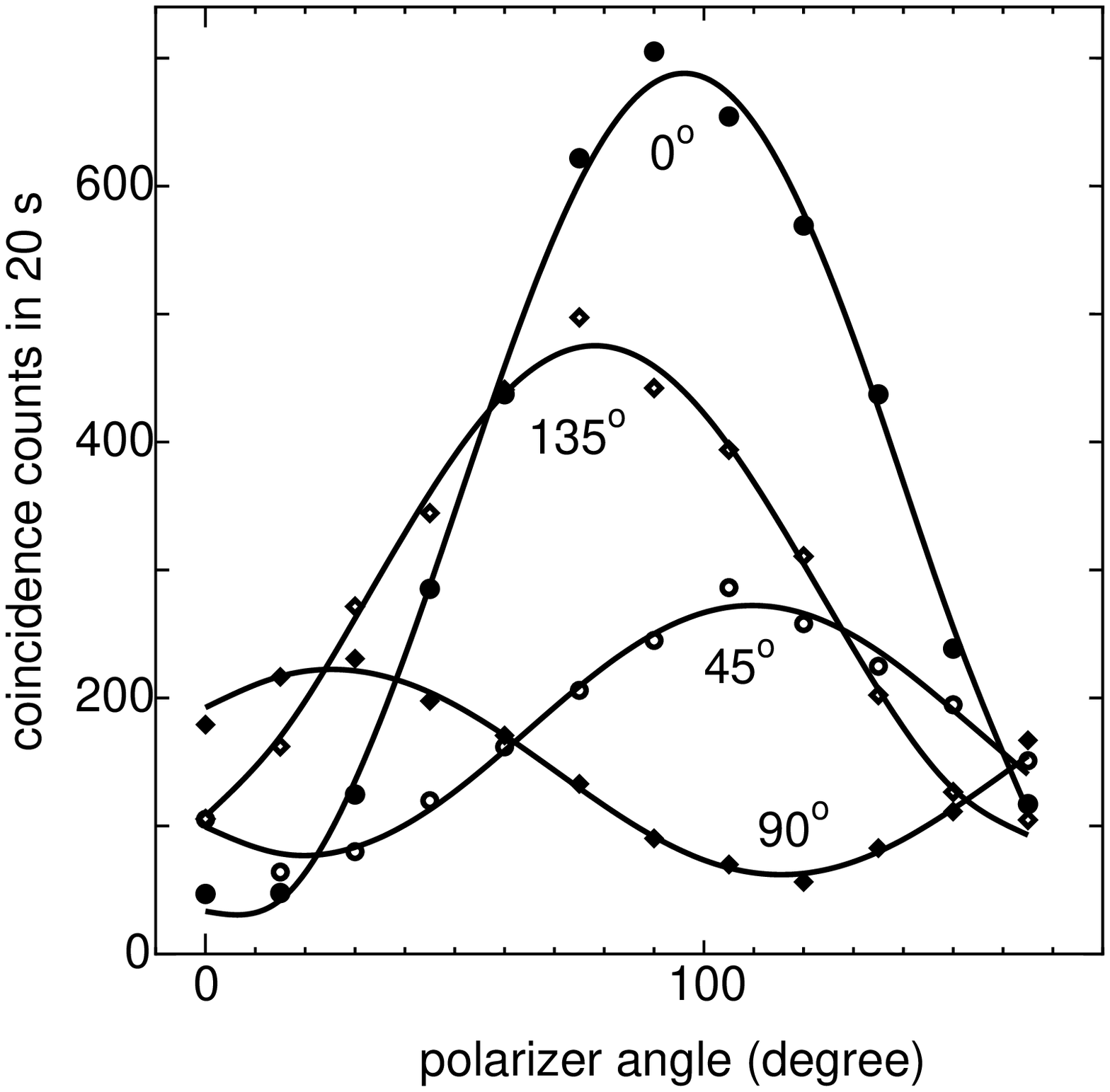}
(b)
    \caption{Measured coincidence counts of polarization correlated
     photon pairs. Marks in the
     figure show the results measured by rotating the angle of the
     polarizer POL2, when the polarizer POL1 was fixed to $0^o$ (filled
     circle), $45^o$ (open circle), $90^o$ (filled diamond) and
     $135^o$ (open diamond), respectively. Curves show the results
     fitted by a function, $c\left(1+v\cos(\theta+\theta_0)\right)$. (a)
     the diameter of the iris was 12~mm. (b) the diameter of the iris was
     set to be 1~mm.\label{data-0th}}
    \end{center}
   \end{figure}

As is mentioned before, an iris makes SPDC photon pairs to be
entangled in their polarizations. It was confirmed under the condition
when the detector of signal photons was pointed to the zeroth-order
diffraction light from the grating, by measuring the polarization
correlation of the SPDC photon pairs when the diameter of the iris was
12~mm or 1~mm (see Fig.~\ref{data-0th}).
The coincidence counts were measured by rotating
the angle of the polarizer (POL1) for several angles ($0^o,45^o,90^o$
and $135^o$) of the polarizer (POL2). The zero degree of the angle was
defined by the hardware origin of the motorized continuous rotation
stage, on which the polarizer was fixed. When the angle was zero degree,
the polarizer axis was nearly parallel to the vertical axis.
The counted data were fitted by a function,
$c\left(1+v\cos(\theta+\theta_0)\right)$, with fitting parameters of
$c,v$ and $\theta_0$.

% fitting
The result shows that the SPDC photon pairs were not entangled when the
diameter of the iris was 12~mm, and that the pairs were entangled when
the diameter of the iris was 1~mm. It means that SPDC photon pairs were
entangled by spatially selecting the crossing point of the signal light
cone and the idler light cone.

\begin{figure}[htbp]
    \begin{center}
     \includegraphics[scale=0.4]{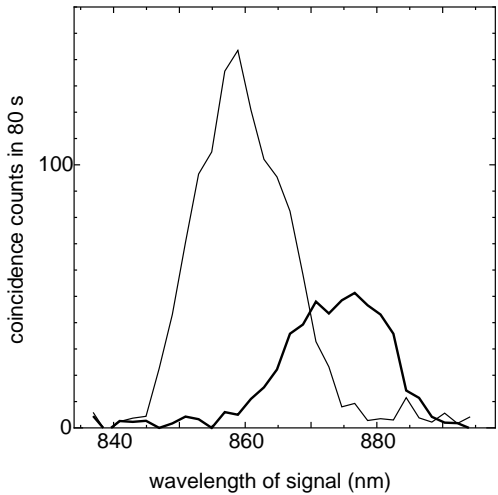}
    \end{center}
    \caption{The idler wavelength dependency of the coincidence counts
    of $|H\rangle_s|V\rangle_i$ (thick curve) and
    $|V\rangle_s|H\rangle_i$ (thin curve). \label{data-spectrum-both}}
   \end{figure}

The main point of this article is to show that the SPDC photon pairs are
entangled after being resolved by their frequencies, by measuring the
polarization correlations of frequency resolved SPDC photon pairs.
To compare with the theoretical
simulations, the parameter $f$ in Eq.~(\ref{eq-biphoton}) must be
measured. The parameter can be calculated from the coincidence counting
rates of $|H\rangle_s|V\rangle_i$ and $|V\rangle_s|H\rangle_i$.
The signal wavelength dependency of the coincidence counting
rate of $|H\rangle_s|V\rangle_i$ was measured by rotating the grating
around the vertical axis crossing the incident point of the signal beam,
setting the angles of the polarizers POL1 and POL2 were set to transmit
vertically polarized idler photons and horizontally polarized signal
photons, respectively.
Then, the idler wavelength dependency of the coincidence counting
rate of $|V\rangle_s|H\rangle_i$ was measured by rotating the grating
around the vertical axis crossing the incident point of the idler beam.
This was performed by letting horizontally polarized idler photons and
vertically polarized signal photons transmit through POL1 and POL2 by
selecting their angles, respectively (see
Fig.~\ref{data-spectrum-both}).
The results show that the coincidence counting rate of
$|H\rangle_s|V\rangle_i$ was about three times larger than that of
$|V\rangle_s|H\rangle_i$ when the signal wavelength was 866~nm, and both
rates were nearly balanced when the signal wavelength was 870~nm.
%The differences in intensity and spectral position and shape is due to
%slight displacement of the locations of detectors.

\begin{figure}[htbp]
    \begin{center}
     \includegraphics[scale=0.4]{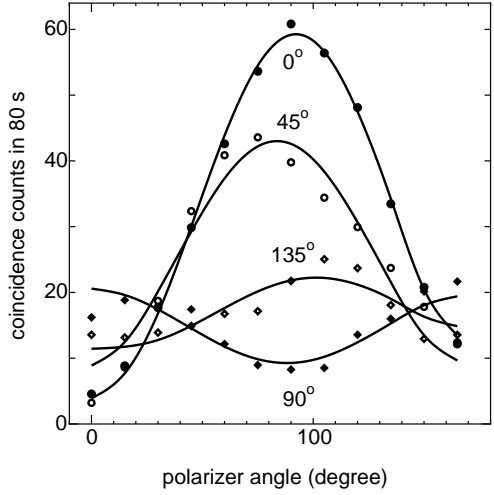}
    \caption{Measured coincidence counts of polarization correlated
     photon pairs when the coincidence
     counts of $|H\rangle_s|V\rangle_i$ and $|V\rangle_s|H\rangle_i$ were
     not balanced. Marks in the figure show the results measured by
     rotating the angle of the polarizer POL2, when the polarizer POL1
     was fixed to $0^o$ (filled circles), $45^o$ (open circles), $90^o$
     (filled diamonds) and $135^o$ (open diamonds), respectively. Curves
     show the results fitted by a function,
     $c\left(1+v\cos(\theta+\theta_0)\right)$ with fitting parameters of
     $c,v$ and $\theta_0$.\label{data-unbalance}}
    \end{center}
   \end{figure}

Experiment was performed under two conditions as follows. First, the
angle of the grating was set to diffract 866~nm signal photons
to the detector, and the diameter of the iris was set to be
1~mm. Figure~\ref{data-unbalance} shows the measured polarization
correlation. The fact that $\Theta_i(45^o)$ and $\Theta_i(135^o)$ were
shifted from $\Theta_i(0^o)$ by $10^o (<45^o )$, indicates that
even though the SPDC photon pairs were entangled but they were not
maximally entangled. As is mentioned in the
section of theory, the angular rotation was less than $45^o$ if
$f\neq1$. Actually, Fig.~\ref{data-spectrum-both} shows that
$f\approx1.73$ when the signal wavelength was 866~nm.
% fitting

\begin{figure}[htbp]
    \begin{center}
     \includegraphics[scale=0.4]{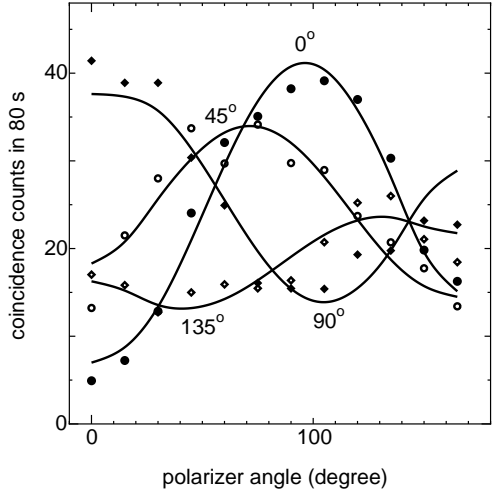}
    \caption{Measured coincidence counts of polarization correlated
     photon pairs when the coincidence
     counts of $|H\rangle_s|V\rangle_i$ and $|V\rangle_s|H\rangle_i$ were
     balanced. Marks in the figure show the results measured by rotating
     the angle of the polarizer POL2, when the polarizer POL1 was fixed
     to $0^o$ (filled circles), $45^o$ (open circles), $90^o$ (filled
     diamonds) and $135^o$ (open diamonds), respectively. Curves show the
     results fitted by a function,
     $c\left(1+v\cos(\theta+\theta_0)\right)$ with fitting parameters of
     $c,v$ and $\theta_0$.\label{data-balance}}
    \end{center}
   \end{figure}

Second, the angle of the grating was set to couple 870~nm signal photons to
the detector, and the diameter of the iris was set to
1~mm. Figure~\ref{data-balance} shows the measured result of the
polarization correlation. The shift of $\Theta_i$ was larger than the
case when the detector was coupled to 866~nm signal photons. It is
because $f\approx1$ in this case. Actually, the coincidence counting
rate of $|H\rangle_s|V\rangle_i$ was nearly equal to the rate of
$|V\rangle_s|H\rangle_i$ (see Fig.~\ref{data-spectrum-both}).
% fitting
The shift of $\Theta_i$ has increased from about $10^o$ to
$30^o$. However the shift was still smaller than $45^o$, and the
visibility of the polarization correlation was less than one.
It means the polarization entanglement was not
maximized. It can be improved by compensating walk-off effect and group
velocity dispersion in the non-linear crystal. The compensation can be
performed by inserting a half-wave plate and a non-linear crystal of
which thickness is half of the crystal which was used to generate SPDC
photon pairs~\cite{pol-entangle-kwiat}.

There are two points which are necessary to implement an ideal
source of WDM-QKD. One point is that the photon pairs should be
broadband in the spectrum and entangled in frequency and
polarization. The generation of the broadband frequency-entangled photon
pairs has been succeeded in~\cite{yab-SPDC-spectroscopy}, and the photon
pairs can be entangled in polarization with a small modification in the
setup as mentioned in~\cite{yab-SPDC-spectroscopy}. The other point is
that the coincidence counts of $|H\rangle_s|V\rangle_i$ and
$|V\rangle_s|H\rangle_i$ are balanced in all over the spectral range of
the SPDC photon pairs. It can be accomplished by using optical
components whose efficiencies have no polarization dependency. Actually,
the transmittance/reflectance of the real beam splitter and the diffraction
efficiency of the grating have polarization dependency in the real
experimental setup used in the present study, resulting in the
unbalanced coincidence counts of $|H\rangle_s|V\rangle_i$ and
$|V\rangle_s|H\rangle_i$. Proper improvement of these two points,
the SPDC photon pairs will be performed in the implementation of WDM-QKD,
which can send more information than the normal QKD, because each
spectral channel in WDM-QKD can send same amount of information which
the normal QKD can transfer.

\section{Conclusion}

In conclusion, it was verified that the frequency resolved SPDC photon
pairs were polarization-entangled in appropriate configurations.
When the iris transmits only photons passing through the crossing point
of the signal light cone and the idler light cone, the SPDC photon pairs
were entangled in polarization. The shift of $\Theta_i$ was $10^o$ when
the coincidence counts of $|H\rangle_s|V\rangle_i$ and
$|V\rangle_s|H\rangle_i$ were unbalanced, and the shift has increased to
$30^o$ when they were balanced. However the observed shift was smaller
than $45^o$. It means that the polarization entanglement of the photon
pairs was not maximized, because of the walk-off effect and group velocity
dispersion in the non-linear
crystal. These can be improved by utilizing a half-wave plate and a
half-thickness non-linear crystal. Requirements for the real
applications of WDM-QKD have been discussed. We thank Drs. Haibo Wang
and Yongmin Li and Mr. Tomoyuki Horikiri for their valuable discussion.

%\bibliography{../bib/myrefs,../bib/quantum}

\end{document}